\documentclass[10pt,conference]{IEEEtran}
\IEEEoverridecommandlockouts
\usepackage{cite}
\usepackage{algorithm,algorithmic}
\usepackage{amsmath,amssymb,amsfonts}
\usepackage{algorithmic}
\usepackage{graphicx}
\usepackage{textcomp}
\usepackage{xcolor}
\usepackage{subfigure}
\usepackage{diagbox}
\usepackage[usestackEOL]{stackengine}
\usepackage{subfigure}
\usepackage{multirow}
\usepackage{ctable}
\usepackage{url}
\def\BibTeX{{\rm B\kern-.05em{\sc i\kern-.025em b}\kern-.08em
    T\kern-.1667em\lower.7ex\hbox{E}\kern-.125emX}}
\begin{document}

\title{Design and Evaluation of a Rack-Scale Disaggregated Memory Architecture For Data Centers}

\author{\IEEEauthorblockN{Amit Puri, John Jose, Tamarapalli Venkatesh}
\IEEEauthorblockA{\textit{Dept. of CSE, IIT Guwahati, Assam, India} \\
email: \{amitpuri, john.jose, t.venkat\}@iitg.ac.in}}

\maketitle
\begin{abstract}
Memory disaggregation is being considered as a strong alternative to traditional architecture to deal with the memory under-utilization in data centers. Disaggregated memory can adapt to dynamically changing memory requirements for the data center applications like data analytics, big data, etc., that require in-memory processing. However, such systems can face high remote memory access latency due to the interconnect speeds. In this paper, we explore a rack-scale disaggregated memory architecture and discuss the various design aspects. We design a  trace-driven simulator that combines an event-based interconnect and a  cycle-accurate memory simulator to evaluate the performance of disaggregated memory system at the rack scale. Our study shows that not only the interconnect but the contention in the remote memory queues also adds significantly to remote memory access latency. We introduces a memory allocation policy to  reduce the latency compared to the conventional policies. We conduct experiments using various benchmarks with diverse memory access patterns. Our study shows encouraging results towards the rack-scale memory disaggregation and acceptable average memory access latency.
\end{abstract}


\setlength\columnsep{0.42164cm}
\section{Introduction}
With high-end server class multi-processors like Xeon Phi and AMD's EPYC, the compute capability in servers has improved dramatically, with ability to run multiple applications simultaneously. However, the typical workloads in high-performance computing (HPC) facilities and cloud data centers such as big data analytics, and machine learning applications, fall short of server memory due to under-utilization and the memory capacity wall~\cite{10.1145/1555754.1555789}. Due to improper use of on-board memory, memory gets stranded as small fragments within each server node increasing the total cost of ownership~\cite{10.1145/2391229.2391236}. Disaggregation of memory resources allows a modular approach to manage memory in a fine-grained manner, where memory does not need to be on the same board as the processor. It allows independent upgrade of memory and increases the data center hardware refresh cycle time~\cite{esaasa15}.In this paper, we study a rack-scale system with partial memory disaggregation where each compute node has some local memory to fulfill the primary requirements, while most of the application memory requirements get fulfilled by the remote memory. The remote memory is managed in the form of multiple remote memory pools within the same rack and is allocated to the compute nodes on-demand.

Disaggregated memory introduces several design challenges. First, the placement of remote memory outside the board should be such that multiple compute nodes can access remote memory simultaneously, without significant congestion. Second, how should the remote memory address space be exposed to avoid system-level bottlenecks with little overhead. Another requirement is a centralized memory manager to take care of remote memory allocation that should also balance load across memory pools. Different types of remote memory access require the support of different interconnect design and protocols. Cache-based access requires a memory binding fabric support for faster access \cite{7446090,7753261}, whereas accessing remote memory in larger chunks requires support for Remote direct memory access (RDMA) \cite{9415599}. We propose a two-level remote memory allocation mechanism, one at the compute node level and other at the global memory manager. Our study shows that different memory allocation methods can impact the performance in pool-based remote memory.The contributions of this work are as follows:
\begin{itemize}
\item We explore a rack-scale design for memory disaggregation and discuss the design space.
\item We identify the major factors impacting the remote memory access latency and propose cost-effective memory allocation policies that also perform load-balancing to get rid of tail latency.
\item We evaluate the performance of proposed memory allocation policy on diverse workloads to show the overall impact of memory disaggregation.
\end{itemize}
\section{Related Work and Motivation}
Earlier designs proposed memory disaggregation at rack-scale for traditional server nodes \cite{6522317,10.1145/2658260.2658262,10.1145/1851476.1851495}. Infiniswap \cite{10.5555/3154630.3154683}, and FARM \cite{179767} presented optimizations for virtually disaggregated systems to utilize RDMA access to remote memory and leverage free memory in other servers. Lim. et al. \cite{10.1145/1555754.1555789,6168955} present a general-purpose physical disaggregated memory design, where memory blades are connected to the compute nodes through PCIe buses. Scale-out NUMA \cite{10.1145/2541940.2541965} presented an on-chip hardware block to provide a low latency interface between the processor and remote memory. Venice \cite{7446090} and DEOI \cite{7753261} also explored similar on-chip modules for remote memory access with separate channels for fine-grained and paged access to remote memory. Recently, a consortium of hardware industry leaders released protocol standards for a similar on-chip memory coherent interconnect, Gen-Z that includes a switch and a pooled memory subsystem \cite{9289390}. Komareddy et al. proposed a shared memory approach for pooled memory with a single remote address space to all the compute nodes \cite{10.1145/3286475.3286480}. On the other hand, a large class of data center applications can fulfill their CPU demands from the computing power available within a single system and only require shared excess to memory in a few instances. Under such a scenario, multiple compute nodes rarely require shared access to the remote memory. Instead, retaining coherency inside a single domain will prevent coherency traffic and reduce significant overhead. Our work leverages the non-coherent use of pooled memory systems to allocate remote memory. Furthermore, evaluation of disaggregated memory systems has earlier been done at a small scale either in a virtualized environment \cite{10.1145/2391229.2391236}, or with a host OS \cite{10.1145/3124680.3124731} by adding fixed network latency and fixed division of address space into local and remote. 

\section{Rack Scale Design}
Our approach to memory disaggregation considers only rack-scale remote memory access, because going beyond a rack will increase the latency further due to network latency.
\subsection{Pooled Memory Management}
The compute nodes will not only depend on remote memory for most of their memory allocation, but the design should also support extensive memory requests from these nodes in a time-bound manner. Remote memory on facing contention in its memory queues adds significant tail latency, which makes it compulsory to have smaller pools of memory and more communication points to get high overall memory bandwidth. There is no consensus yet on the number of memory pools within a rack, as disaggregated designs are still in an experimental stage. Our base design includes remote memory in the form of multiple remote pools. Our experiments in later sections will establish a correlation between the compute node's workload demands and the number of memory pools.

\subsection{Remote Memory Organization}
The remote memory can be made transparent to all compute nodes, where the operating system at the node can directly allocate memory from any part. However, all the nodes should have a consistent view through a global memory manager. This approach has scalability issue due to significant coherency traffic to the shared memory. Another approach is to provide mapped access to the remote memory, where any remote memory page exclusively belongs to a single node. The global memory manager can reserve the remote memory in larger chunks (a few megabytes) without any bottleneck. We chose distributed access to memory in our design.

\begin{figure}[t]
\centering
\subfigure[RMAC integrated On-Chip ]
    {
        \includegraphics[width=3.6cm,height=2.25cm]{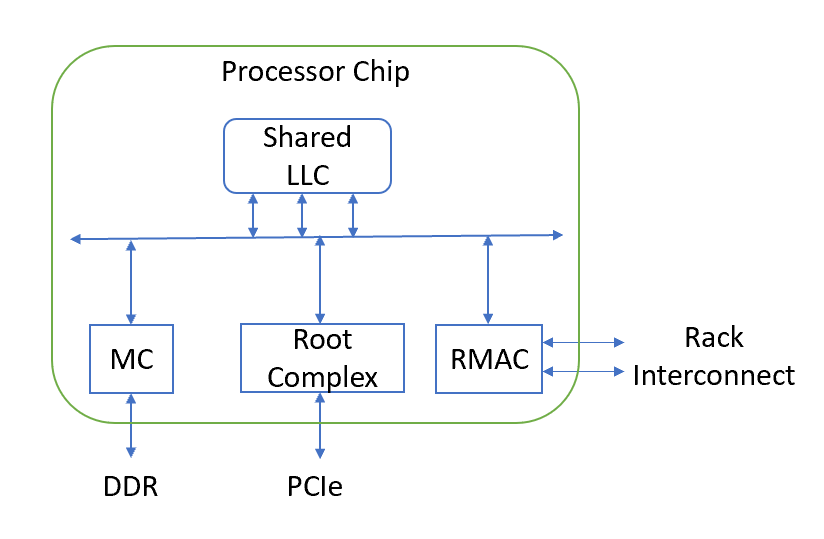}
        \label{fig3a}
    }
\subfigure[Forwarding using RMAC]
    {
        \includegraphics[width=3.6cm,height=2.4cm]{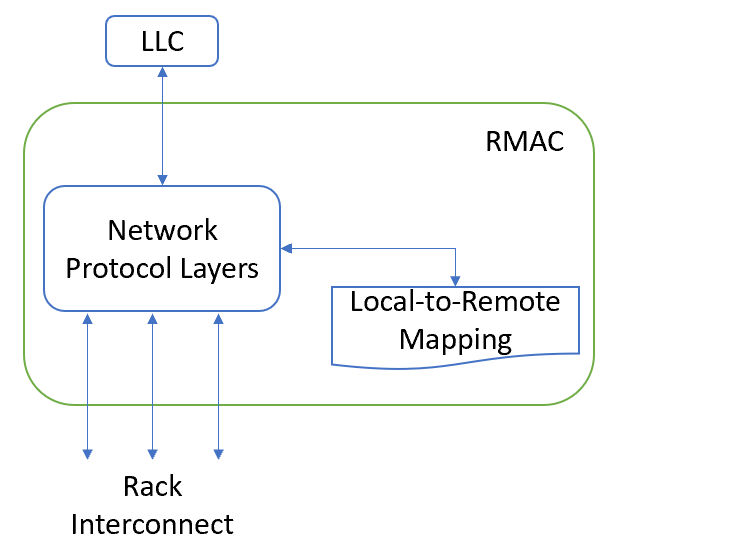}
        \label{fig3b}
    }
\caption{Remote memory access with RMAC.}
\label{fig_4-5}
\end{figure}
\subsection{Interconnect Requirements}
Even though the fast network switches had substantially decreased the network latency, a large part of network overhead is due to the node's deep protocol stack, slow I/O buses, and protocol conversion while offloading requests off the chip. An on-chip network interface such as a remote memory access controller (RMAC) shown in Fig. \ref{fig3a} holds the key for future data centers that join the remote memory resources for cache-based load/store. RMAC is an addressable device that also takes care of the bookkeeping mechanism required for routing cache misses toward the appropriate memory pool. Such on-chip interconnects that enable quick remote memory access have been explored in the past \cite{7446090,7753261,10.1145/2541940.2541965,9137193,5238684} and are a good match for data center applications. As shown in Fig. \ref{fig3b},  RMAC forwards the last-level cache (LLC) miss requests that belong to remote memory and implements a lightweight network protocol on the hardware. On the other hand, coarse grain page access can be implemented as a DMA-like channel over the same interface that works with a user or kernel space daemon to monitor hot remote memory pages and occasionally bring them to local memory. RDMA interconnects such as RoCE \cite{5238675} and InfiniBand \cite{4154093} that allows one-sided access to remote memory are already in use in present data centers \cite{10.5555/3026877.3026897}.
\subsection{Global Memory Manager}
A global memory manager manages all the remote memory within a single rack. Whenever an application falls short of the local memory, it causes page fault requesting the global memory manager to allocate a chunk from one of the memory pools, which forms an extended local memory address space on compute node. Linux allows online up-gradation of system memory using memory hot-plug service, which is exposed to the OS page allocator once initialized. It is also essential to allocate the memory in smaller chunks to have more granular control over remote memory. If the allocation size is too small, the mapping table will grow huge and incurs significant search latency. If it is too big, memory will be under-utilized like in traditional servers, which makes remote memory reclamation challenging, requiring large amounts of data migration. The global memory manager can be hosted at the ToR switch and will maintain the memory tables for the allocated and free remote memory. Once it reserves a remote memory chunk, the manager sends the chunk details to the requesting node to add suitable local-to-remote mappings at RMAC for address translation.


\begin{figure*}[htbp]
\centering
\subfigure[Average memory access latency]
    {
        \includegraphics[width=4.2cm,height=3.3cm]{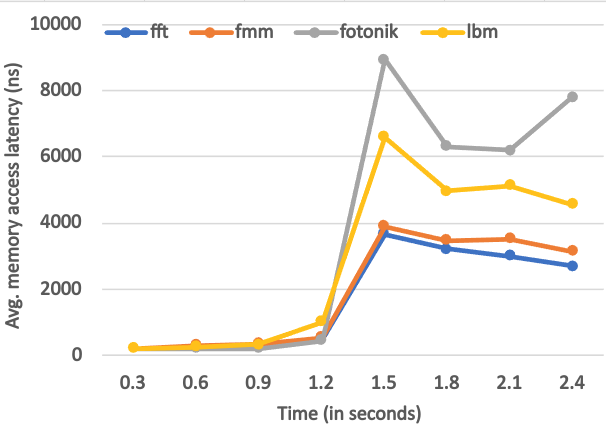}
        \label{g16}
    }
\subfigure[Remote memory access latency]
    {
        \includegraphics[width=4.2cm,height=3.3cm]{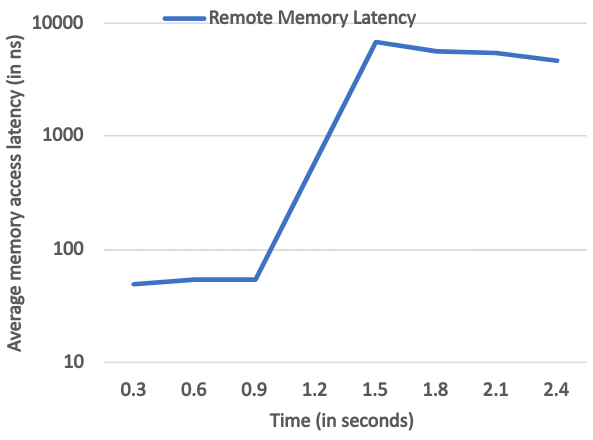}
        \label{g17}
    }
\subfigure[Latency distribution]
{
    \includegraphics[width=4.2cm,height=3.3cm]{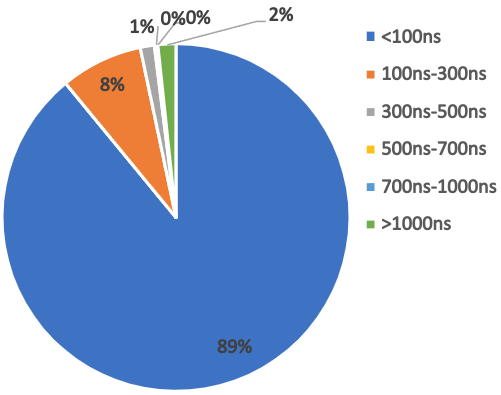}
    \label{g18}
}
\subfigure[Access variation in memory pools]
{
    \includegraphics[width=4.2cm,height=3.3cm]{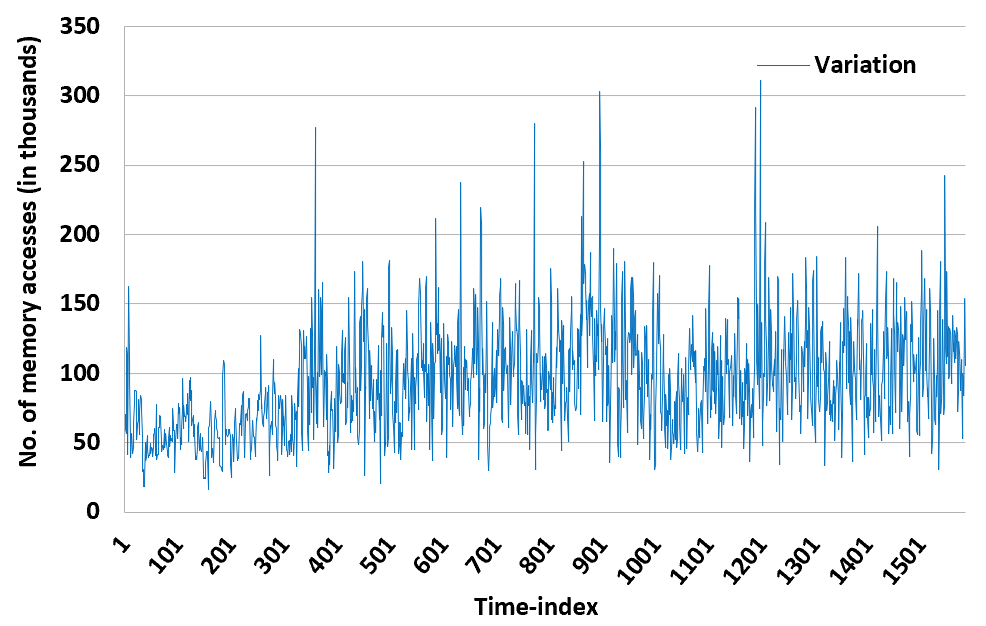}
    \label{g19}
}
\caption{Random pool selection with alternate local-remote page allocation}
\label{g16_g17_g18_g19}
\end{figure*}
\section{Rack-Scale Memory Allocation}
Memory allocation in disaggregated memory systems is two-fold. Firstly, the page allocation policy on compute nodes must decide when to start utilizing the remote memory. A node has an option to use \textit{Local memory first}, or it can use an \textit{Alternative Local-Remote} approach for allocating consecutive pages. With the first one, it will initially enjoy the benefits of fast memory but suffer a sudden slow down once the local memory is exhausted. Many applications tend to go through a start-up phase and do not benefit from this scheme. The other option will allow better average memory access latency for an extended period but does not make full use of fast local memory. Secondly, the global memory manager warrants a pool-selection policy for allocating a chunk of memory. Although network latency is the major hurdle in remote memory access, our study shows that pool-selection policy significantly impacts the average memory access time on remote pools. In a pooled remote memory system, a memory pool connects to one of the switch links. Without load-balancing, few memory pools that get more requests will face tail latency due to congestion at switch buffers.
\subsection{Random Pool Selection}
We initially analyze workload \textit{WL-Mix} (explained in section-V) with a random pool selection, for which the global manager randomly selects a pool for every chunk allocation request by a compute node OS (4MB allocation size) and the pages are allocated alternatively in local and remote memory. As shown in fig \ref{g16}, the average memory access latency for each benchmark has exceeded to microseconds, which is way beyond the manageable limit at which an application can execute normally. The only significant factor for this high latency is the contention in remote memory queues, which is the consequence of blindly allocating a random remote pool and using it for the application's memory requirements. The same can be observed in fig \ref{g17} that shows only the average remote memory latency, excluding the network delays. We probed all the memory accesses to find that \textit{2\%} of the memory accesses (only remote), with latency of \textit{1000ns} or more, led to high average latency (shown in fig \ref{g18}). In fig \ref{g19}, we show the variation in the number of memory accesses across different pools for every \textit{1.5} million-cycles. The variation is calculated by subtracting the pool with maximum and minimum memory access during that period. The large variation shows the imbalance of memory access among the pools, causing high tail latency.

\subsection{Smart-Idle Pool Selection}
This policy performs optimal memory pool selection in two different steps. The first step selects a small subset of memory pools from all the available pools based on the recent memory access traffic. The rationale is that the memory pools with less current traffic are least likely to face contention soon and can be selected currently for more memory allocation. The second step will finally select a pool from the subset with the least amount of already allocated memory. The reason behind this is to balance the amount of memory allocation equally among pools. Another reason is that even if a memory pool is currently facing less traffic, it can still suddenly face more memory requests from the previously allocated memory if that pool has been allocated more memory in the past. So the choice of pool with the least allocated memory is less likely to face such sudden accesses. 
To implement this, we use the global memory manager hosted at the rack switch that keeps track of total memory accesses to each remote pool. It only requires a small number of 32 or 64-byte counters. We use a \textit{window-based} mechanism to determine the traffic by measuring an access factor (\textit{Af}) of each pool at the start of every window. A \textit{window} is the duration between allocating two consecutive remote memory chunks. Memory accesses of four recent windows are maintained to determine the activity of each remote pool since it reflects the most recent status of the memory traffic. The access factors (\textit{Af}) is calculated as described in \eqref{eq1}, and \textit{MemAccCount} in \eqref{eq2} refers to the total number of memory accesses to a memory pool in a window. More weightage is given to memory access count in the recent window compared to older windows to get most recent status. A lower value of \textit{Af} indicates that a pool has faced less memory traffic recently and can be selected for the next chunk allocation.

\begin{equation}
\label{eq1}
Af_{win(n+1)} =  M_{aC}(New) + M_{aC}(Old)/3
\end{equation}
\begin{equation}
\label{eq2}
M_{aC}(New)  = MemAccCount_{win(n)}
\end{equation}
\begin{equation}
\label{eq3}
M_{aC}(Old) =   \sum_{z=n-1}^{n-3}MemAccCount_{win(z)}
\end{equation}

The smart-idle allocation makes sure to choose a lesser active pool while also evenly distributing the memory chunks across pools. Assuming the total number of memory pools to be \textit{n}, smart-idle policy will initially select a set of pools with set-size \textit{m}, where \textit{m} is calculated as: \begin{math}
m = Ceil [log2 (n)].\end{math} Then will finally choose a pool with lowest allocated memory.

\section{Experiment Methodology and Results}
We simulate all the main memory accesses for an application, combined to represent many nodes running inside a rack which are finally processed to simulate the interconnect and memory. Once the traces are ready, the task is to simulate each node's main memory accesses at local or remote memory based on the reference address. The front-end uses Intel's PIN \cite{10.1145/1064978.1065034} platform to perform binary instrumentation for the application analysis. Our tool is based on a \textit{Allcache} pin-tool that performs a functional simulation of TLB and cache hierarchy. The base tool is modified to support multi-threaded trace collection and give an approximate timing for a TLB/cache miss, over which instrumentation was performed at instruction-level granularity by collecting LLC misses in the same way as in \cite{Alachiotis2019}. A combined trace is sorted by time-stamp of merged cache-misses from all the cores. The LLC misses eventually give the main memory accesses that also preserve the multi-threaded nature of the application, where each record has a virtual address of the LLC miss, its time-stamp, thread-id, and read/write access type. The virtual addresses in the trace are translated by simulating a memory management unit, that tracks pages in local and remote memory. For every page fault, it allocates a new page in local or remote memory. A global memory manager is also simulated to serve remote memory allocation requests from nodes.

The simulation for interconnect occurs as a set of discrete events where an event is one CPU cycle. A queue-based mechanism simulates latency for NIC as well as rack-level interconnect. We use finite-sized queues for NIC and rack switch ports with a back-pressure congestion control policy and add appropriate queuing delays to the waiting requests once the queues get full. Further, propagation delays and transmission delays are added to each packet according to wire-length and link speeds respectively. Each remote memory access is sent in the form of network packets, for which packing and unpacking time is added appropriately. A switch arbitrator selects the ready packets from virtual queues at input ports, avoiding starvation and head-of-line blocking. Finally, we use DRAMSim2 \cite{5732229} to simulate the main memory for which multiple instances were deployed, each for local memory at compute nodes and remote memory in memory pools. Sorted memory access from a multi-core front-end facilitates memory simulations in an environment representing multi-threaded execution. 

\renewcommand{\arraystretch}{1.1}

\begin{table}[htbp]
\caption{Benchmarks}
\begin{center}
\begin{tabular}{ccccc}
\specialrule{.1em}{.05em}{.05em}
\textbf{Benchmark} & \textbf{Cache} & \textbf{RAM Accesses} & \textbf{Footprint} & \textbf{Label}\\
\textbf{Name} & \textbf{Miss-Rate} & \textbf{(in Millions)} & \textbf{(in GB)} &\\
\hline
lbm\_s & 12.49\% & 45.47 & 2.7 & \multirow{4}{*}{\textit{WL-Mix}}\\
fotonik3d\_s & 5.96\% & 11.92 & 0.57 &\\
fft   & 4.16\% & 15.81 & 1.06 &\\
fmm   & 3.42\% & 12.5 & 3.20 &\\
\specialrule{.1em}{.05em}{.05em}
\end{tabular}
\label{tab4}
\end{center}
\end{table}

\begin{table}[htbp]
\caption{Simulation Parameters}
\begin{center}
\begin{tabular}{cc}
\specialrule{.1em}{.05em}{.05em}
\textbf{Element} & \textbf{Parameter}\\
\hline
CPU & 1.2GHz, 8-core\\ 
ITLB & 128 Entries, 8-Way ITLB, 60-cycle\\ 
DTLB & 64-Enteries, 4-Way DTLB, 60-cycle \\
Cache Size & 32KB(L1-I/D), 256KB(L2), 16MB(L3)\\
Cache Associativity & 8-Way(L1), 4-Way(L2), 16-Way(L3)\\
Cache Latency & 4-Cycle(L1), 12-Cycle(L2), 41-Cycle(L3)\\
Cache Type & Write-Back/Write-Allocate, Round-Robin, 64B\\
Memory & 256MB Per node, 32GB per Pool\\
Switch & 128x 100Gbps, 132mb Buffer, 20ns delay\\
RMAC (NIC) & 100Gbps, 10ns Delay\\
Packet-Size & 64B request, 128B response, 25ns Packet-Prep\\
\specialrule{.1em}{.05em}{.05em}
\end{tabular}
\label{tab3}
\end{center}
\end{table}


\begin{figure*}[htbp]
\centering
\subfigure[Round-Robin pool selection]
    {
        \includegraphics[width=5.8cm,height=3.2cm]{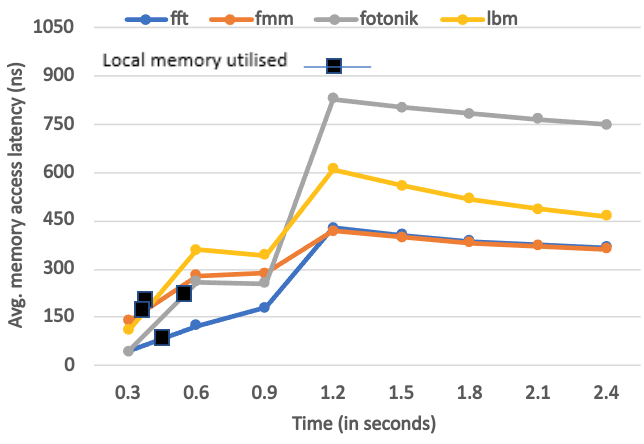}
        \label{g8}
    }
\subfigure[Smart-Idle pool selection]
    {
        \includegraphics[width=5.8cm,height=3.2cm]{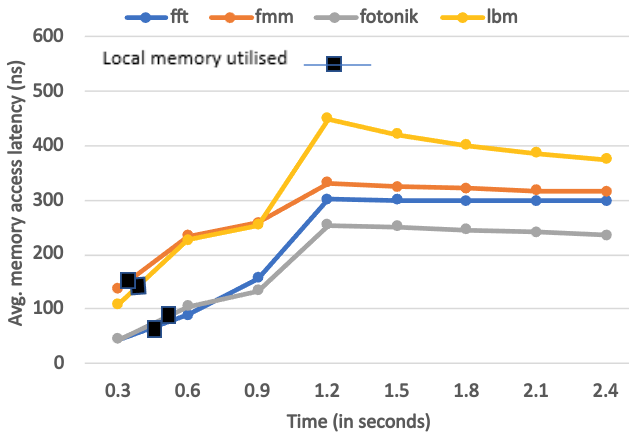}
        \label{g9}
    }
\subfigure[Average remote memory latency]
{
    \includegraphics[width=5.5cm,height=3.2cm]{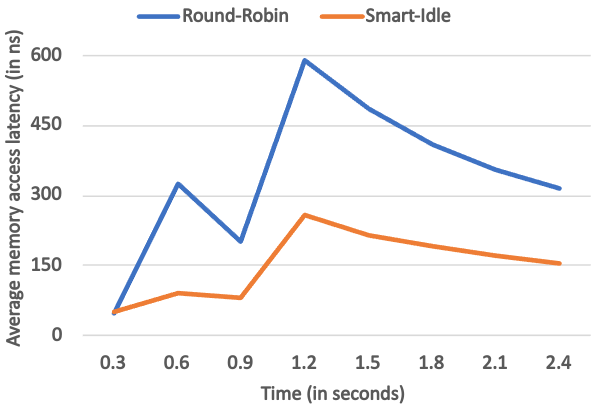}
    \label{g10}
}
\caption{Average memory access latency with \textit{Local-First Allocation}}
\label{g8_g9_g10}
\end{figure*}

\begin{figure*}[htbp]
\centering
\subfigure[Round-Robin pool selection]
    {
        \includegraphics[width=5.8cm,height=3.2cm]{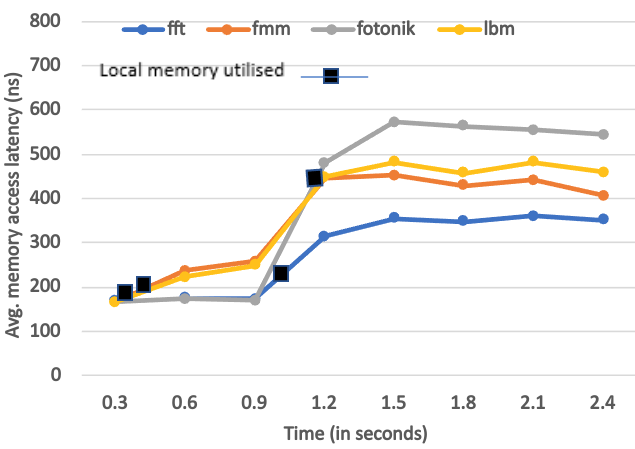}
        \label{g11}
    }
\subfigure[Smart-Idle pool selection]
    {
        \includegraphics[width=5.8cm,height=3.2cm]{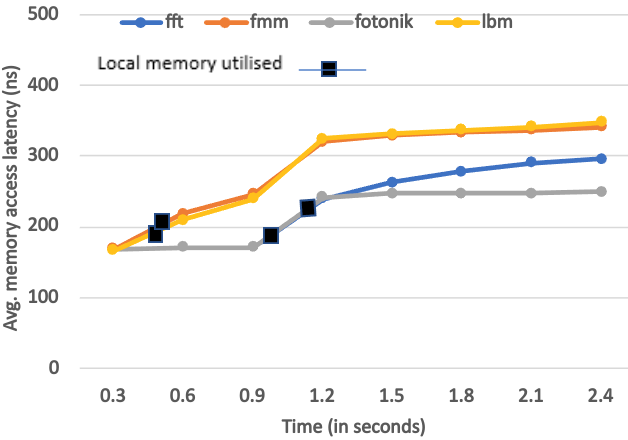}
        \label{g12}
    }
\subfigure[Average remote memory latency]
{
    \includegraphics[width=5.5cm,height=3.2cm]{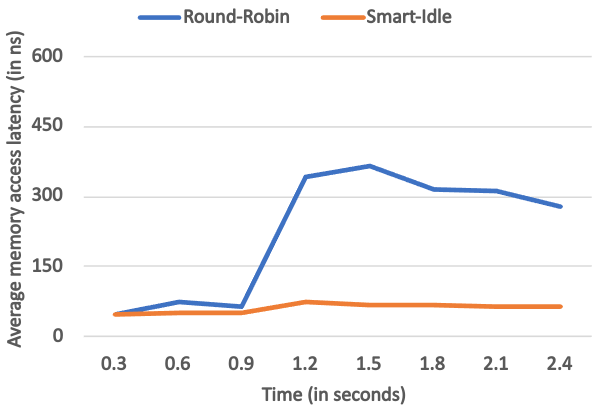}
    \label{g13}
}
\caption{Average memory access latency with \textit{Alternate Local-Remote Allocation}}
\label{g11_g12_g13}
\end{figure*}

\begin{figure*}[htbp]
\centering
\subfigure[Remote Memory Latency Distribution]
    {
        \includegraphics[width=8.75cm,height=3.4cm]{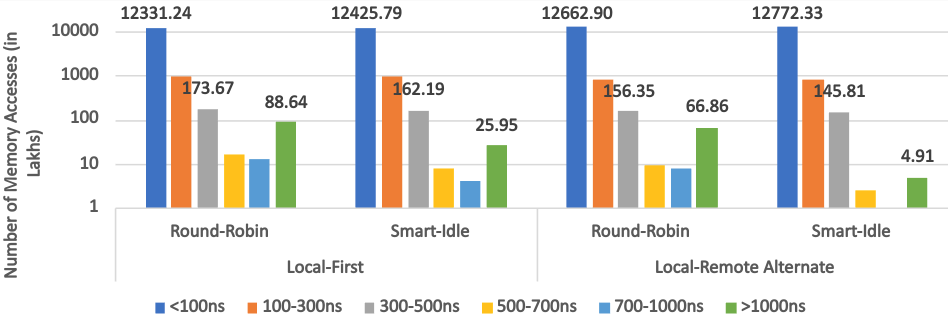}
        \label{g14}
    }
\subfigure[Latency breakdown]
    {
        \includegraphics[width=8.75cm,height=3.4cm]{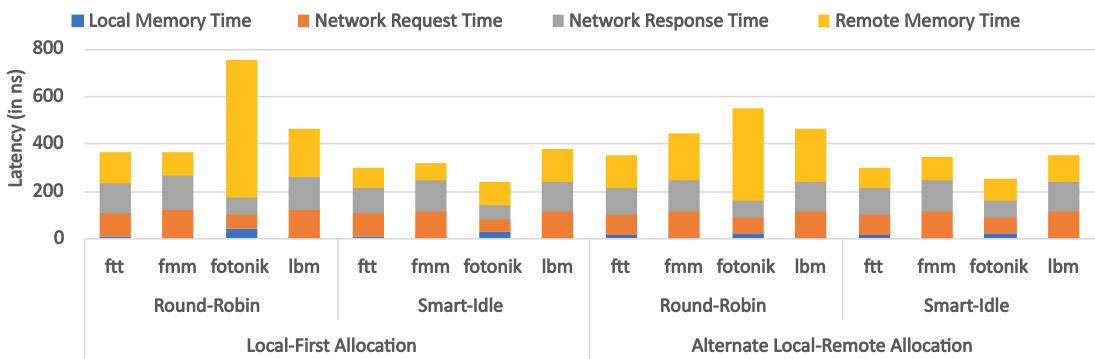}
        \label{g15}
    }
\caption{Distribution of remote memory accesses based on access times and Latency breakdown for Local/Remote/Network time}
\label{g14_g15}
\end{figure*}

We choose 4-multi-threaded benchmarks, shown in table \ref{tab4}. Each benchmark have large variation in total number of memory access made during the simulation time and represents the heterogeneous workload of data center servers. The workload mix \textit{WL-Mix} are deployed for rack-scale experimentation with 64-compute nodes and 6-memory pools, where one workload is deployed on 16 nodes each having 256MB of local memory. We intentionally kept the number of memory pools on the lesser side, as our motive here is to test the memory pools' maximum-bandwidth limits for a high workload scenario. In table \ref{tab3}, we sum up all the system parameters used for the simulations. We perform the experiments over both \textit{local-first} and \textit{alternate local-remote} page allocation policies that are run with Round-Robin and Smart-Idle pool selection.

We first discuss \textit{local-first} page allocation for round-robin pool selection in Fig. \ref{g8}, which shows the cumulative average memory access latency at different simulation points, where black marks represent the time when no more local memory is left. Although the results show a substantial decrease in the average access latency compared to the random pool selection, it is still high, especially for \textit{lbm} and \textit{fotonik}. This is because both these benchmarks send the most memory access to remote memory. The average latency of \textit{fft} and \textit{fmm} is not good either to maintain sufficient application speed. On the other hand, as shown in fig. \ref{g9}, smart-idle improves the average latency to a significant margin compared to round-robin pool selection. None of the benchmarks face serious contention at remote memory queues except after epoch4 to some extent, despite which smart-idle kept the average latency down throughout. Even with local-first allocation, latency only increased gradually for all of the benchmarks, which was the result of proper load-balancing. Fig. \ref{g10} shows the cumulative average memory access latency at the memory pools (without including network delays) for both round-robin and smart-idle pool selection, which are completely in sync with the above results. Due to a sudden burst of memory requests in between, the round-robin could not perfectly handle balancing these requests across memory pools. However, with smart-idle allocation, chunk allocation was done so that memory accesses would be divided almost equally across multiple pools, which is why it gives a better result. We next measure the performance with \textit{alternate local-remote} allocation, shown in fig. \ref{g11_g12_g13}. As expected, there is no sudden burst of remote memory accesses, and we observe a gradual increase in latency for all benchmarks once the local memory is finished. Surprisingly, both round-robin and smart-idle perform relatively better than they performed with \textit{local first}. It does not show much impact of not having exclusive access to the fast local memory initially and even though \textit{lbm} and \textit{fmm} have a large memory footprint, they are still able to achieve good enough average memory latency. Overall, We saw the same trend being followed here also, where smart-idle performs better than the round-robin. However, the latency difference was less this time. These results show that the overall average memory latency is most optimized with \textit{alternate local-remote} page allocation combined with smart-idle pool selection, shown in fig. \ref{g12}.

Further analysis of the completion time of all the remote memory accesses is shown in Fig. \ref{g14}. This latency only includes a memory request's time at the remote memory before it completes the memory access. Different colored bars here represent the number of memory accesses completed in each category based on its access latency. While the latency with random pool selection was higher, round-robin pool selection brought it down. There are a large number of memory requests beyond 500ns (bars in yellow, light blue, and green). Round-robin policy is not sufficient to balance the memory traffic equally across different pools. The smart-idle pool selection combined with local-remote alternate page allocation is better than a simple round-robin to reduce the tail latency. The graph shows that very few accesses take more than 500ns to complete. We show the overall latency breakdown for all the memory accesses in Fig. \ref{g15}. Smart-idle suffers lesser network delays than round-robin as memory request packets are distributed equally across multiple links connecting the memory pools. However, there is a big variation in average remote memory access time for all the benchmarks through different policies.

\section{Conclusion and Future Work}
In this paper, we explored rack-scale memory disaggregated systems, that provide more flexibility in memory utilization but come with some overheads. Remote memory access delay is another aspect we looked into through our experiments. High contention in the memory queues of remote memory became a dominating factor in most cases when pool selection was done through conventional policies. The proposed smart-idle pool selection policy evenly distributes the memory traffic load among all the memory pools to counter the high tail latency and provides a much more balanced combined average access latency to local and remote memory. Further, it will be interesting to see the improvement with optimization like remote prefetching or remote page migration to local memory. Disaggregated memory systems will require strategically chosen pages to be migrated from shared remote memory for each node. Considering such approaches to mask the remote memory latency would be part of our future work.
\bibliographystyle{IEEEtran}
\bibliography{HPCC}

\end{document}